\documentclass{article}
\usepackage{graphicx}
\usepackage{amsmath}
\usepackage{amsfonts}
\usepackage{amssymb}
\newtheorem{theorem}{Theorem}

\newtheorem{definition}[theorem]{Definition}

\begin{document}

\title{Positivity and Integrability \\{\small (Mathematical Physics at the FU-Berlin)}\\{\small To Michael Karowski and Robert Schrader on the occasion of their 65th} {\small birthday}}
\author{Bert Schroer\\CBPF, Rua Dr. Xavier Sigaud 150 \\22290-180 Rio de Janeiro, Brazil\\and Institut fuer Theoretische Physik der FU Berlin, Germany}
\date{November 2005}
\maketitle
\begin{abstract}
Based on past contributions by Robert Schrader and Michael Karowski I review
the problem of existence of interacting quantum field theory and present
recent ideas and results on rigorous constructions.
\end{abstract}

\section{Historical remarks}

The title of this essay is identical to that of a small conference at the
FU-Berlin in honor of Michael Karowski and Robert Schrader at the occasion of
their sixty-fifth birthday. The history of mathematical physics and quantum
field theory at the FU-Berlin, a university which was founded at the beginning
of the cold war, is to a good part characterized by ''positivity and
integrability'' \cite{con}.

Both of my colleagues joined the FU theory group in the first half of the 70s
shortly after I moved there. Robert Schrader arrived after his important
contribution \cite{O-S} to the birth of Euclidean field theory whose proper
mathematical formulation he initiated together with Konrad Osterwalder while
working at Harvard university under the guidance of Arthur Jaffe; Michael
Karowski came from Hamburg where he finished his thesis under Harry Lehmann.
Whereas Robert, after his arrival in Berlin, was still in the midst of
finishing up the work with he begun with Konrad Osterwalder at Harvard
\cite{O-S2}, Michael was looking for new challenging post-doc problems outside
his thesis work. At that time Dashen, Hasslacher and Neveu \cite{D-H-N} (DHN)
had just published their observations on the conjectured exactness of the
quasiclassical particle spectrum of certain 2-dimensional models. There were
some theoretical indications \cite{S-T-W} and numerical checks \cite{Berg}
pointing to a purely elastic S-matrix in those apparently integrable models
which were strongly suggestive of an explanation in the (at that time already
discredited) S-matrix bootstrap setting, but now within a more special context
of factorizing elastic S-matrices. It was already clear at that time that such
a structural property is only consistent in low spacetime dimensions. In the
hands of Michael Karowski and a group of enthusiastic collaborators (B. Berg,
H-J. Thun, T.T. Truong, P. Weisz) the S-matrix principles based on crossing,
analyticity and unitarity behind these (at that time still experimental
mathematical) observations were adapted to two-dimensional purely elastic
2-particle scattering. The findings were published in a joint paper \cite{K}
which together with a second paper on the subject of formfactors \cite{K-W}
associated with those factorizing bootstrap S-matrices became the analytic
basis for systematic model constructions of quantum field theories based on
the bootstrap-formfactor program. The aspect of integrability of these models
was verified by constructing a complete set of conserved currents. During this
time Robert exploited the new Euclidean framework in order to obtain a
constructive control of models. For this purpose he had to use a more
restrictive sufficient criterion which limited the class of models to those
whose short distance behavior is close to that of free fields.which turned out
to be possible in 2-dim. QFT.

All these developments took place in a city which was the most eastern outpost
of the western world and for this reason played the role of a show-window of
capitalism\footnote{The capitalism of the cold war was the traditional one
i.e. of a different kind as today. These days it is often referred to as the
``Rhenish capitalism'' in German publications in order to distinguish it from
the much more US version..} and liberty which together with its geographic
isolation contributed to its well-maintained infrastructure and high quality
of life with poverty and social deprivation being virtually unknown. The
relative isolation of the city was moderated by a lush funding for inviting
guests and there was a very large number of internationally known visitors far
beyond the list of international collaborators which each of us attracted
after joining the FU faculty.

From those early papers of Michael Karowski and co-workers it became clear
that some of the far out speculative conjectures of the Californian
(Chew-Stapp...) S-matix bootstrap ideas on uniqueness \footnote{The S-matrix
bootstrap at that time was propagandized as a new theory of everything (apart
from gravity). As we all know this was neither the first (it was preceded by
Heisenberg's ``nonlinear spinor theory'') neither the last time. But whereas
the old attempts ended in a natural death, the more modern versions are still
(not unlike the legendary flying dutchman) circeling over our heads in search
of a physical landing place.} were without foundation; to the contrary, far
from being a unique characterization of a a theory of everything (TOE), the
two-dimensional scheme of factorizing S-matrices led to a rich classification
of two-dimensional QFT which contained besides the mentioned DHN models many
others of physical interest. These new non-perturbative methods attracted a
lot of attention and the classification of factorizing S-matrices and the
construction of associated integrable models of QFT has remained a fascinating
area of QFT ever since. 

Through my scientific contacts with Jorge Andre Swieca in Brazil (we both were
research associates under Rudolf Haag at the university of Illinois) this line
of research took roots at the USP in Sao Paulo and other Brazilian
universities and via students of Swieca and Koeberle it led to the formation
of a whole group of researchers (E. Abdalla, F. Alcaraz, V. Kurak, E.
Marino,..) and also influenced others who nowadays are important members of
the Brazilian theoretical physics community.

Besides these two groups which vigorously pursued these new ideas about an
S-matrix based construction of low dimensional field theories, there was
another independent line with similar aims but stronger emphasis on algebraic
structures; this was pursued at the Landau Institute in Moscow (the
Zamolodchikov brothers and others \cite{Za}) as well as at the Steklov
Institute in St. Petersburg (L. Faddeev, F. Smirnov and others \cite{Fa}%
\cite{Smir}). There were many scientific exchanges especially with the people
from the Steklov institute.

Unfortunately the FU work (this applies in some lesser degree also to
Schrader's contributions) did not generate the interests which it deserved
within the quantum field theoretical establishment in Germany
(Lehmann\footnote{There is some irony in the fact that Lehmann at the end of
his life became actively interested in low dimensional theories. But at that
time his influence on particle physics in Germany was already declining.}%
,Symanzik, Zimmermann and Haag), and without this support this became an
uphill struggle for the FU group. Several of the highly gifted members of the
young FU research group had to end their academic careers; when the
recognition of their achievements outside of Germany had a positive feedback
within it was already too late\footnote{One outstanding younger member of the
group was H. J. Thun whose joint work with Sidney Coleman \cite{C-T} on the
nature of higher poles in the S-matrix became a standard reference. Despite
his impressive start there was no place for his academic carrier in Germany.}.

After the fall of the Berlin wall the situation with respect to fundamental
QFT research in Berlin worsened. The physics department of the Humboldt
University was restructured solely by ``Wessies'' (to use the colloquial
Berlin slang of those days which has survived up to the present); the
home-grown community in QFT had no say and was not asked for advice.
Considering the importance of historical continuity in particle theory and the
intellectual damage caused by political interference, it is not surprising
that the consequences of those negative influences have left their mark and
will become even more evident after two of the last FU-QFT innovators go into
retirement. Outside attempts to fill fundamental quantum field theoretical
research with the passing flow of theoretical fashions did not work.

But this article is not primarily an essay of past achievements of two of my
colleagues, nor about the history of QFT and the ups ans downs of mathematical
physics in Berlin. I rather prefer to demonstrate the relevance of their past
innovations by convincing the reader that the legacy of their ideas constitute
still an important part of the ongoing scientific dialogue.

In order to be able to do this, I first have to recall some of the
pre-electronic conceptual advances which got lost or failed to get passed in
the proper way to the younger generation. A good illustration of such loss of
knowledge as a result of oversimplifications and distortions caused by
globalized fashions is the fate suffered by the \textit{Euclidean method.} The
work of Osterwalder and Schrader and prior contributors started with a very
subtle and powerful problematization of what is behind the so-called Wick
rotation. These days I sometimes find myself attending talks in which the
speaker did not bother to explain whether he is in the setting of real time
QFT or in the Euclidean setting; a related question during such a talk usually
raises the speakers eyebrows because to him in his intellectual state of
innocence it is evident that one can pass from one to the other without being
bogged down by the need of justification. The rapid rate in which fundamental
knowledge gets lost or substituted by universal phrases is one of the symptoms
of a deep crisis in particle physics. \ 

Indeed the Euclidean theory associated with certain families of real time QFTs
is a subject whose subtle and restrictive nature has been lost in many
contemporary publications as a result of the ``banalization'' of the Wick
rotation (for some pertinent critical remarks referred to in \cite{Re}). The
mere presence of analyticity linking real with imaginary (Euclidean) time
without checking the validity of the subtle reflection positivity (which is
necessary\ to derive the real time spacelike commutativity as well as the
Hilbert space structure) is not of much physical use.

In times of lack of guidance from experiments the most reasonable strategy is
to press ahead with the intrinsic logic of the existing framework, using the
strong guidance of the past principles and concepts rather than paying too
much attention to the formalism which was used to for their implementation. In
a previous particle physics crisis, namely that of the ultraviolet
divergencies of QFT, it was precisely this attitude and not the many wild
speculations in the decade before renormalization theory, which finally led to
the amazing progress; in fact renormalization theory was probably the most
\textit{conservative affirmation of the underlying causality and spectral
principles} of Jordan's ``Quantelung der Wellenfelder''. What was however
radically different was their new mathematical and conceptual implementation.

Although this impressive progress made QFT what it is today, namely the most
successful physical theory of all times up to this date, it is still suffering
from one defect which sets it apart from any other area of theoretical
physics. Whereas in other areas the construction of models preceded the
presentation of a setting of axioms (which extract the shared principles
underlying the explicit constructions), things unfortunately did not work this
way in QFT. The reason was that the perturbative constructions (unlike say in
mechanics, astronomy and quantum mechanics) did not come with mathematical
assertions concerning their convergence and estimates of errors. These aspects
were not only missing in Feynman's perturbative (and any differently
formulated) approach, but it became increasingly clear that all these series
were at best only asymptotically converging (which unfortunately is a property
which does not reveal anything about the mathematical existence).

This led the birth of an axiomatic framework\footnote{This terminology has
often been misunderstood. It has a completely different connotation than say
axiomatics of mechanics or thermodynamics, since it results from the
realization that it is much harder to do a credible computation for a concrete
model than it is to understand joint structural properties of a whole class of
models (as long as the question of existence is ignored).} which was followed
by constructive QFT. Measured in terms of the complexity of the problem, the
results and the methods by which they were obtained are impressive
\cite{Gl-Ja}. There was also a certain amount of elegance which clearly came
from a very clever use of Euclideanization and/or algebraic properties.
Conceptually these methods followed closely the physical ideas underlying
renormalized perturbation theory. As in the perturbative approach the main
objects to be controlled are correlation functions of pointlike fields, with
free fields and their Fock space still playing an important auxiliary role. An
important technical step was to establish a measure-theoretical interpretation
to the interaction-polynomial relativ to the free field measure..As a result
of methodical limitations it was virtually impossible to go beyond the very
restrictive superrenormalizability requirement and to incorporate real life
models (as e.g. the Standard Model). For reasons of a certain imbalance
between an unwieldy mathematical formalism and the few (and mostly anyhow
expected) results besides the control of existenc, the \textit{constructive
approach} did not enter standard textbooks but rather remained in the form of
reviews and monographies; expert cynics even sometimes referred to it (in
particular by H. Lehmann) as ``destructive QFT''. Most practitioners of QFT do
not mention this problem or raise their students with the palliative advice
that since QFT is not credible for very short distances, the existence problem
is somewhat academic; but unfortunately without giving the slightest hint how
in physics (i.e. outside of politics) problems can be solved by enlarging
them. Recently there has been some renewed interest in a variant of this
method which is based on the hope that some progress on the functional
analytic control \cite{Schle} of time-dependent Hamiltonian problems may
extend the mathematical range (of a more modern algebraic formulation) of the
Bogoliubov S-operator approach beyond what had been already achieved in
\cite{Wre}.

For a number of years I have entertained the idea that one is struggeling here
against a birth-defect of QFT whose effects can be only removed by a some
radical conceptional engineering. I am referring here to the that
\textit{classical paralellism} called Lagrangian quantization by which Pascual
Jordan found the ``Quantelung der Wellenfelder''\footnote{His peer Max Born
with Heisenberg's support limited Jordan's obsession with quantizing also
structures beyond mechanics by banning his calculations to the last section of
the joint work. Darrigol \cite{Dar} reports that when Jordan received
Schroedinger's results he already had what was later called a second quantized
version. For his radical viewpoint it was apparently sufficient that a
structure could be fitted into a classical framework and not whether it was
actually part of classical physics.}. What is most amazing is the fact that
only two years after his discovery he apparently became worried about this
kind of quantization not being really intrinsic to quantum physics (at that
time, shortly after computing the Jordan-Pauli commutatot function, he also
could have been already aware that pointlike quantum fields as obtained from
Lagrangian quantization are singular objects besides lacking intrinsicness).
Although we do not know his precise motivation, as the main plenary speaker at
the first post QFT big international conference 1929 in Kharkov \cite{Kha}
(probably the last international sympsium in which German was the conference
language) he pleaded to look for a new access to QFT which avoids such
``classical crutches'' (klassische Kruecken) but without proposing a way to
implement this idea. Apart from Wigner's isolated representaion-theoretical
approach to relativistic particles in 1939 (which only in the late 50's became
known to a broader public through the work of Wightman and Haag), the first
entirely intrinsic setting which avoided field coordinatizations
(non-intrinsic generators of spacetime-indexed algebras) was the algebraic
approach by Haag (with some mathematical ideas concerning operator algebras
taken from Irving Segal), later known as the Haag-Kastler approach. But in
some sense the baby was thrown out with the bath water because the conceptual
precision was not matched by any calculational implementation. This was at
least the situation before the modular theory of operators algebras was
incorporated in order to place local quantum physics on a more constructive
course. At this point Karowski's contribution (and more generally the
meanwhile extensive literature on the classification and construction of
factorizing model) enters; these explicit and nontrivial model constructions
represent presently the most valuable theoretical laboratory for the new
constructive ideas based on modular theory; they have led to new results
concerning the proof of existence of certain strictly renormalizable models to
which we will return later. These new developments suggest that it rather
improbable to win a prize for the solution of gauge theory as an isolated
subject without revolutionizing the whole of local quantum physics.

The main motivations for writing this essay is to point out that both the
issue of Euclideanization and the bootstrap-formfactor approach are both by no
means mature closed subjects; rather they are illustrative examples which
shows that QFT despite its age is still very far from its closure.

The article is organized as follows. The next section reviews some of the
ideas which led to the Osterwalder-Schrader results and their role in
strengthening the constructive approach to QFT. In the same section I also
sketch the more recent framework of modular localization. Its analyticity
aspects derive from the domain properties of a certain unbounded operator
which characterizes localization aspects of operator subalgebra; unlike the
Bargmann-Hall-Wightman analyticity of correlation functions of pointlike
fields (operator-valued distributions) which constitute an important aspect of
the Euclidean approach, the Tomita-Takesaki modular theory does not refer to
individual operators but rather encodes joint properties of operators which
are members of an operator algebra. Euclideanization aspects of modular theory
are presented in section 4. The third section sketches the notion of
integrability in QFT which is synonemous with factorization of the S-matrix,
but in order to maintain a unified conceptual line I deviate from the
historical path and present factorizing models in the modular setting. In the
last section I list some open problems related to the theme of this essay.

\section{Positivity and Euclideanization}

After the rehabilitation of the divergence-ridden QFT in the form of
renormalized perturbative quantum electrodynamics, and after the subsequent
conceptual advances in the understanding of the particle-field relation
through scattering theory \cite{LSZ}\cite{Haag}, the idea gained ground that
the importance of QFT for particle physics can be significantly enlarged by
understanding more about its model-independent structural properties beyond
perturbation theory. The first such general setting was that by Wightman
\cite{St-Wi}. In harmony with the increasing importance of analytic properties
which entered the setting of scattering theory through the particle physics
adaptation of the optical Kramers-Kronig dispersion relations, special
emphasis was placed on the study of analytically continued correlation
functions. These investigations started from the \textit{positive energy
momentum} spectrum (expressing the presence of a stable ground state, the
vacuum) and the \textit{Lorentz-covariant transformation} properties as well
as \textit{locality} (in the form of (anti)commutators vanishing for spacelike
distances). This, led to an extension of the original tube domain resulting
from the positive energy-momentum spectrum to the Bargmann-Hall-Wightman
region of analyticity and its extension by locality. The continued correlation
functions turned out to be analytic and uni-valued in the resulting ``permuted
extended tube'' region \cite{St-Wi}.

Already before these mathematical results the Euclidean region, which resulted
by letting the time component be pure imaginary, attracted the attention of
Schwinger since in his formalism it led to some computational simplification.
The next step was taken by Symanzik \cite{Sy} who observed from his functional
integral manipulations that the analytic continuation to imaginary times (Wick
rotation) for bosonic theories highlighted a positivity which was well-known
from the continuous setting of statistical mechanics. Nelson \cite{Ne}
succeeded to remove the somewhat formal aspects and achieved a perfect
placement into a mathematically rigorous setting of an autonomous stochastic
Euclidean field theory (EFT) in which the most important conceptual structure
was the Markov property. Guerra \cite{Gu} applied this new setting to control
the vacuum energy coming from certain polynomial interactions in bosonic
2-dimensional QFT and emphasized its usefulness in establishing the existence
of the thermodynamic limit. This Euclidean field formalism was limited to
fields with a canonical short distance behavior; but even in this limited
setting \textit{composite fields} with worse short distance behavior permit no
natural incorporation into this probabilistic setting. The limitation was
intimately related to the property of Nelson-Symanzik positivity, which
basically is the kind of positivity which Schwinger functions should obey if
they were to describe a (continuous) stochastic classical mechanics.

The breakthrough for the understanding of Euclideanization of the general
situation in QFT was achieved in the work of Osterwalder and Schrader
\cite{O-S}. They had to sacrifice the stochastic interpretation of EFT which
was then substituted by a certain reflection positivity condition as well a
growth condition on n-point Schwinger functions for n $\rightarrow\infty$. If
one is less ambitious and only asks for a sufficient condition on Schwinger
functions, one obtains a formulation which turns out to be quite useful for
controlling the existence for certain low-dimensional QFTs.

One reason why the constructive control of higher dimensional QFTs present a
serious obstacle is that the reflection positivity does not harmonize well
with the idea of (Euclidean invariant) ultraviolet cutoffs. For this reason
one encounters serious difficulties with ultraviolet cutoffs in a functional
integral setting; in general one does not even know whether such a cutoff is
consistent with the quantum theory setting; not to mention all the other
requirements as e.g. cluster properties, asymptotic scattering limits etc.
which one needs to maintain the physical interpretation of a theory. Of course
cutoff versions are strictly auxiliary constructs and as such may violate such
properties, but then the control of the cutoff-removal becomes a hairy
problem. On the other hand it is possible to formulate the process of O-S
Euclideanization by starting from the more algebraic setting of AQFT which
avoids the use of (necessarily singular) point-like field coordinatisations;
in this case one has problems to specify concrete interactions. \ A
formulation of the algebraic approach in the lattice setting for which the
concepts and their mathematical implementation of the O-S euclidianization
allow a very simple presentation can be found in \cite{Ba-Fr}.

One property within the Nelson-Symanzik setting which turned out to be
extremely useful in controlling the removal of infrared regularizations
(thermodynamic limit) is the Euclidean spacetime duality. This
\textit{Nelson-Symanzik duality} is suggested by the formal use of the
Feynman-Kac Euclidean functional integral representation. Let us consider
thermal correlation functions at inverse temperature $\beta$ for a
2-dimensional enclosed in a periodic box (rather interval). The KMS condition
for the correlation functions at imaginary times reduces to a $\beta
$-periodicity property. Since the Euclidean functional representation treats
space and time on equal footing, the duality under a change of $x$ and $t_{E}$
accompanied by an exchange of the box- with the thermal- periodicity is
obvious. The mathematical physics derivation of this result can be found in a
recent paper \cite{Ge-Ja}. In the last section we will use this property as an
analogy of the chiral \textit{temperature duality}. A model-independent
systematic adaptation of the O-S Euclideanization to the imaginary time
thermal setting can be found in a recent review paper \cite{Bi-Fr}.

In the remainder of this section I will recall the modular localization
setting for the convenience of the reader. This is a preparatory step for the
content of the last section. The salient properties of the modular aspects of
QFT can be summarized as follows \cite{M-S-Y}.

\begin{itemize}
\item  Modular localization is an adaptation of the modular Tomita-Takesaki
theory in the setting of operator algebras. The analytic properties are not
associated to local covariant fields but rather to the operator algebra
$\mathcal{A(O)}$ which is associated with smeared fields if one limits the
test function supports to a fixed spacetime region $\mathcal{O}$. Modular
theory \cite{Su} is based on the idea that one learns a lot about operator
algebras by studying the unbounded antilinear and (as it turns out) closed
operator $S$ defined as%
\begin{equation}
SA\Omega=A^{\ast}\Omega,\text{ }A\in\mathcal{A}%
\end{equation}
where $\Omega$ is a cyclic (i.e. $\mathcal{A}\Omega$ is dense in $H$) and
separating (there is no nontrivial $A\in\mathcal{A}$ which annihilates
$\Omega$). Interesting properties arise from its polar decomposition which is
traditionally written as
\begin{equation}
S=J\Delta^{\frac{1}{2}}%
\end{equation}
The resulting unbounded positive operator $\Delta$ generates via its
one-parametric unitary group $\Delta^{it}$ a modular automorphism group of
$\mathcal{A}$ and the ``angular'' part $J$\textbf{, }the so-called Tomita
conjugation, is an antiunitary involution which maps the operator algebra into
its commutant $\mathcal{A}^{\prime}$%
\begin{equation}
\sigma_{t}(\mathcal{A})\equiv\Delta^{it}\mathcal{A}\Delta^{-it}\subset
\mathcal{A,}\text{ }J\mathcal{A}J=\mathcal{A}^{\prime}%
\end{equation}
where I used a condensed notation using $\mathcal{A}$ as a short hand for its
individual operators $A\in\mathcal{A}$. The crucial property of the Tomita $S$
which is behind all this algebraic richness is the fact that $S$ is
''transparent'' in the sense that $domS=ranS=dom\Delta^{\frac{1}{2}}%
,\,S^{2}=1$ on $domS.$ I am not aware of the existence of such unusual (not
even in Reed-Simon) operators outside modular theory. This theory begins to
unfold its magic power within QFT\footnote{Actually the constructive power of
the modular approach only began to unfold after a seminal paper by Borchers
\cite{CPT} which led to a flurry of additional remarks \cite{Froe}\cite{mod}
and finally gave rise to the theory of modular inclusions and modular
intersections \cite{Wie}\cite{Bo}.} once one realizes (as was first done by
Bisognano and Wichmann \cite{Bi-Wi}) that not only any pair ($\mathcal{A(O)}%
,\Omega=vacuum$) with a nontrivial spacelike disjoint $\mathcal{O}^{\prime}$
is ``standard'' in the sense of modular theory but even more: for the standard
pair ($\mathcal{A(W)}$,$\Omega$) with $\mathcal{W}$ a wedge region, the
modular group acts as the unique W-preserving Lorentz-boost and the Tomita
reflection is (up to a rotation which depends on the choice of W) equal to the
physically extremely significant TCP symmetry. Whereas the unitary
$\Delta_{\mathcal{W}}^{it}$ is ``kinematical'' i.e. determined once the
representation theory of the Poincar\'{e} group (the spectrum of particles) is
known, the $J_{W}$ contains profound dynamical information. If we assume that
we are in the standard LSZ setting of scattering theory\footnote{The LSZ
asymptotic convergence of Heisenberg operators towards free (incoming or
outgoing) particle operators is guarantied by spacelike locality and the
assumption of gaps which separate the one-particle massive state from the
continuum \cite{Haag}.} then the $J$ of an interacting theory is connected by
its interaction-free asymptotic counterpart $J_{0}$ through the scattering
matrix
\begin{equation}
J=S_{scat}J_{0}%
\end{equation}
i.e. whereas in the interaction-free case the modular data for the wedge
algebra are constructed in terms of the relevant representation of the
Poincar\'{e} group, the presence of interactions enriches the modular theory
of wedge algebras through the S-matrix. For models for which the bootstrap
construction of their S-matrix can be separated from the construction of their
fields (the factorizing models of the next section) the knowledge of the
modular data can be used for their explicit construction. The guiding ides is
that knowing the modular data for the wedge algebra uniquely fixes the modular
operators for all the other causally complete region. Although there is no
geometro-physical interpretation a la Bisognano-Wichmann for the modular
objects of smaller causally closed spacetime regions (spacelike cones, double
cones), there is no problem in constructing them through the process of
algebraic intersections in terms of wedge algebras (such $\mathcal{O}$ are
necessarily causally closed)%
\begin{equation}
\mathcal{A(O)}=\bigcap_{\mathcal{W}\supset\mathcal{O}}\mathcal{A(W)}%
\end{equation}
The impressive constructive power of this theory already shows up in its
application to the Wigner representation theory of positive energy
representations of the Poincar\'{e} group. The results obtained by combining
Wigner's theory with modular localization go beyond the well-known results of
Weinberg on three counts:

\item  The spatial version of the modular localization method associates
string-like localized fields with Wigner's enigmatic family of massless
infinite spin representations whereas previous attempts at best showed that
these representations are incompatible with point-like localization.

\item  For massless finite helicity representations (photons, gravitons,...)
only the ``field strength'' whose scale dimension increases with the helicity
are pointlike whereas the ``potentials'' with would-be dimension one turn out
to be string-like\footnote{In the strict Heisenberg-Wigner spirit of
observables one rejects unphysical ghosts (which formally make potentials
covariant and pointlike) even though they are only computational catalyzers in
order to obtain observables at the end of the computation.} i.e. pointlike
potentials are incompatible with the Wigner representation theory.

\item  The structural analysis carried out by Buchholz and Fredenhagen on
massive theories with a mass gap suggests strongly that the setting of
interactions may be significantly enlarged by permitting interactions to
possess a string-like localization structure. If one wants to implement this
idea in a perturbative setting one needs massive string-localized free fields.
The application of modular localization leads to \textit{scalar string-like
localized fields }for arbitrary spins.
\end{itemize}

There is another important representation theoretical result from modular
localization for d=1+2 dimensional QFT which according to my best knowledge
cannot be derived by any other method. It is well-known that the (abelian in
this case) spin in this case can have anomalous values which activates the
representation theory of the universal covering $\tilde{P}(3)$ first studied
by Bargmann. Combined with the modular localization theory one is able to
determine the localized subspaces and a ``preemptive'' one-particle version of
a plektonic spin-statistics theorem. In this case the transition from the
Wigner representation to the QFT can however not be done in a functorial way
since there is an inherent vacuum polarization related with nontrivial
braid-group statistics \cite{Mu2}.

It is interesting to compare the setting of modular localization with the O-S
euclideanization. The former also leads to analyticity properties and to
euclidean aspects but in this case they are not coming from
Fourier-transformed support properties and their covariant extension but
rather encode domain properties of unbounded operators\textbf{. } The
connection with analyticity properties and ``Euclideanization'' is through the
KMS-property of $\mathcal{A(O)}$ expectation values in the state implemented
by the vector $\Omega.$ The defining equation for $S$ shows that all vectors
of the form $A\Omega$ are in $dom\Delta^{\frac{1}{2}},$ which means that these
vectors $\Delta^{iz}A\Omega$ are analytic in the strip $-Imz<\frac{1}{2}$.

The most attractive and surprising property of this formalism is the encoding
of geometry of localization in domain properties (and a forteriori in
analyticity) in the sense $S_{\mathcal{O}_{1}}\subset S_{\mathcal{O}_{2}}$ if
$\mathcal{O}_{1}\subset\mathcal{O}_{2}$ and $S_{\mathcal{O}_{1}}\subset
S_{\mathcal{O}_{2}}^{\ast}$ if $\mathcal{O}_{1}><\mathcal{O}_{2}\,\cite{Mu3}.$
Such an intimate relation between domain (and range) properties of unbounded
and geometric localization properties is unique in particle physics and is not
met anywhere else in physics (this probably explains why it is not treated in
books on mathematical-physics methods as Reed-Simon)

In the present stage of development of the modular formalism does not permit a
general classification and constructive control in the presence of
interactions. As the Euclidean formalism it is limited to certain low
dimensional QFT but for quite different reasons. What is interesting is that
the families of low dimensional models covered by the two settings are quite
different. Whereas for the Euclidean approach the limitation is the
traditional one coming from short distance properties, the present limitation
of the modular approach has nothing to do with short distance properties of
pointlike fields but rather is tied to the existence of generators of wedge
algebras $\mathcal{A(W)}$ with simple physical properties, so called tempered
vacuum-polarization-free generators (PFG). Some details will be explained in
the next section. It turns out that this requirement is equivalent to the
S-matrix being factorizing.

The test of existence of a model (which has been defined in terms of its
generators for the wedge-localized algebra) in the modular approach is not
related to its good short distance behavior, but rather consists in the
nontriviality of algebraic intersections.

In the last section I will present a Euclideanization based on modular
localization which shows its analogy to the O-S setting.

\section{Tempered PFG, integrable QFT, factorizing models}

Modular localization offers a surprising way to obtain new insight into field
theoretic integrability and the classification of factorizing models. As
before we assume the existence of isolated one-particle mass-shells which is
sufficient for the validity of scattering theory. The starting point is the
following definition which then leads to two theorems.

\begin{definition}
A vacuum-\textbf{p}olarization-\textbf{f}ree-\textbf{g}enerator (PFG) of a
localized algebra $\mathcal{A(O)}$ is a (generally unbounded) operator
G$^{\#}$ affiliate to this algebra which applied to the vacuum creates a
one-particle state without vacuum polarization admixture%
\begin{align}
G\Omega &  =one-particle\,\,vector\\
G^{\ast}\Omega &  =one-particle\,\,vector
\end{align}
\ 
\end{definition}

It is clear that a (suitably smeared) free field is a PFG for any free field
subalgebra $\mathcal{A(O),}$ but it takes some amount of thinking to see that
the inverse also holds i.e. the existence of a PFG for any causally complete
subwedge region $\mathcal{O}$ implies that $G$ is a smeared free field and
that the superselection-sector generated by $G$ is that of a free field
sector. On the other hand the (in/out) particle creation/annihilation
operators are affiliated to the global algebra. The wedge region is a very
interesting borderline case; the application of modular theory shows that PFGs
in interacting theories do exist in that case i.e. in more intuitive physical
terms: \textit{the wedge localization is the best compromise between particles
and fields in interacting QFTs}. A closer examination reveals that if one
demands that PFGs are tempered in the sense that they have domains which are
stable under spacetime translations, the S-matrix is necessarily purely
elastic \cite{B-B-S}. This in turn reduces the possibilities (excluding
``free'' models with braid group statistics in d=1+2) to d=1+1 dimensional
interacting theories and in that case one indeed has the rich class of
factorizing S-matrices as illustrative examples.

\begin{theorem}
\cite{B-B-S}Tempered PFGs are only consistent with purely elastic S-matrices,
and (excluding statistics beyond Bosons/Fermions), elasticity and
non-triviality are only compatible in d=1+1.
\end{theorem}

The crossing property for formfactors excludes connected elastic 3-particle
contributions \footnote{Private communication by Michael Karowski.
\par
{}} so that the factorizing models actually are the only ones whose wedge
algebras are generated by PFGs. This approach culminates in the recognition
that the generators of the Zamolodchikov-Faddeev algebras are actually the
Fourier transforms of the tempered wedge-localized PFGs; in this way the
computational powerful but hitherto (in the LSZ scattering setting)
conceptually somewhat elusive Z-F operator algebra acquires a physical
spacetime interpretation. Since there are some fine points concerning
wedge-localization in the presence of bound states (associated with certain
S-matrix poles in the physical rapidity strip) I will for simplicity assume
that there are none. For models with a continuous coupling strength (e.g. the
Sine-Gordon model) this is achieved by limiting the numerical value of the
coupling parameter. Let us further assume that the particle is spinless in the
sense of the Lorentz-spin. Then the following theorem holds

\begin{theorem}
Let $Z^{\#}(\theta)$ be scalar Z-F operators i.e.
\begin{align}
Z(\theta)Z^{\ast}(\theta^{\prime})  &  =S_{2}^{{}}(\theta-\theta^{\prime
})Z^{\ast}(\theta^{\prime})Z(\theta)+\delta(\theta-\theta^{\prime}%
)\label{PFG}\\
Z(\theta)Z(\theta^{\prime})  &  =S_{2}^{{}}(\theta^{\prime}-\theta
)Z(\theta^{\prime})Z(\theta)\nonumber\\
\phi(x)  &  =\frac{1}{\sqrt{2\pi}}\int(e^{ip(\theta)x(\chi)}Z(\theta
)+h.c.)d\theta\nonumber
\end{align}
then the non-local fields generate a wedge-localized algebra $\mathcal{A(W)}$
and the coefficient functions $S_{2}$ are the two-particle scattering matrix
contributions of a purely elastic factorizing scattering matrix $S_{scat}$
\end{theorem}

As already stated in the previous section, the algebras for compact spacetime
regions and their pointlike field generators are constructed by forming
intersections of wedge algebras. The relevant calculations are very simple in
the case of interaction-free fields associated with the various families of
positive energy Wigner representations \cite{B-G-L}\cite{M-S-Y}. For the case
at hand they are slightly more involved, reflecting the fact that although the
PFG generators are still on-shell but the creation/annihilation components
have a more complicated algebraic structure. There are two strategies to be
followed depending on what one wants to achieve.

If the aim is to establish the existence of the model in the algebraic
setting, then one must find a structural argument which secures the
nontriviality of intersections of wedge algebras associated to causally
complete spacetime regions. For the case at hand the property of
\textit{modular nuclearity} is sufficient to show nontriviality. There are
some recent interesting partial results in this direction by Lechner
\cite{Le}. Meanwhile there exists a proof which applies to all factorizing
models whose S-matrix depends on a coupling strength such that for weak
coupling there is no bound state \cite{new}.

The underlying physical idea is that the nontriviality is already encoded into
the structure of the wedge algebra generators. In particular in d=1+1, a
property called \textit{modular nuclearity} of the wedge algebra (referring to
the cardinality of phase space degree of freedoms \cite{Bu-Le}) secures the
nontriviality of double cone intersections which is tantamount to the
existence of the model in the framework of local quantum physics. Since the
proof uses the S-matrix in in an essential way it is not surprizing that
certain properties which were extremely hard to obtain in the approach based
on Euclideanization as the condition of asymptotic completeness, are a quite
easy side result of the nontrivial existence arguments.

If on the other hand the aim is to do explicit calculations of observables
beyond the S-matrix, then the determination of the formfactor spaces is the
right direction to follow. In that case one makes a
Glaser-Lehmann-Symanzik-like Ansatz, but instead of expanding the desired
localized Heisenberg operator in terms of incoming creation/annihilation
operators, ones uses the Z-F operators instead%
\begin{equation}
A=\sum\frac{1}{n!}\int_{C}...\int_{C}a_{n}(\theta_{1},...\theta_{n}%
):Z(\theta_{1})...Z(\theta_{n}):d\theta_{1}...d\theta_{n} \label{series}%
\end{equation}
Whereas in the GLZ case the coefficient functions are expressible in terms of
mass-shell projections of retarded functions, the coefficient functions in
(\ref{series}) are connected multiparticle formfactors%
\begin{align}
&  \left\langle \Omega\left|  A\right|  p_{n},..p_{1}\right\rangle ^{in}%
=a_{n}(\theta_{1},...\theta_{n}),\;\ \theta_{n}>\theta_{n-1}>...>\theta_{1}\\
&  ^{out}\left\langle p_{1},..p_{l}\left|  A\right|  p_{n},..p_{l+1}%
\right\rangle _{conn}^{in}=a_{n}(\theta_{1}+i\pi,...\theta_{l}+i\pi
,\theta_{l+1},..\theta_{n})\nonumber
\end{align}
which are boundary values of analytic functions in the rapidity variables. If
we are interested in operators localized in a double cone $A\in\mathcal{A}%
\mathcal{(D)}$ we should look for the relative commutant $\mathcal{A}%
\mathcal{(D)}=\mathcal{A(W)}\cap\mathcal{A(W}_{a}\mathcal{)}^{\prime}$ with
$\mathcal{D=W\cap W}_{a}^{\prime}$ and $\mathcal{W}_{a}$ being the wedge
obtained by spatial shifting $\mathcal{W}$ to the right by a. In terms of the
above Ansatz this means that the looked for $A^{\prime}s$ should commute with
the generators of $\mathcal{A(W}_{a})$ i.e.%
\begin{equation}
\left[  A,U(a)\phi(f)U(a)^{\ast}\right]  =0 \label{com}%
\end{equation}
Since the shifted generators are linear in the Z-F operators and the latter
have rather simple bilinear commutation relations, it is possible to solve the
recursive relation (the kinematical pole relation) iteratively and
characterize the resulting spaces of connected formfactors. Although such
recursive formfactor calculations do not reveal existence since bilinear forms
(formfactors) need not result from operators, the combination with the
previous existence argument would show that the bilinear forms are really
particle matrix elements of genuine operators. In the pointlike limit
$a\rightarrow0$ the equation characterizes the space of formfactors of
pointlike fields. In this case one obtains a basis of this space by invoking
the covariance properties of the Lorentz spin. After splitting off a common
rather complicated factor shared by all connected formfactors, the remaining
freedom is encoded in momentum (rapidity) space polynomial structure which is
similar but more complicated than the analogous structure for formfactors of
Wick polynomials.

For all QFT which are not factorizing (i.e. in particular for higher
dimensional theories) there are no PFGs which generate wedge algebras. In this
case the idea would be to try some kind of perturbation theory. A scenario for
such a construction may look as follows. Starting in zeroth order with
generators which are linear in the incoming creation/annihilation operators,
one defines first order generators of the commutant (localized in the opposite
wedge) by using the perturbative first order S-matrix in $J=J_{0}%
S_{scat}^{(1)}.$ This leads to a first order correction for the coefficient
functions of first order generators of the opposite wedge. The hope would be
that the imbalance in the commuting property with the original generator would
then require a second order correction of $S$ as well as a correction in the
coefficient function and that this, similar to the iterative Epstein-Glaser
approach for pointlike fields could serve as a perturbative analog of the
on-shell bootstrap-formfactor program which bypasses correlation functions of
singular fields and leads to a fresh start for a construction program which is
also capable to handle the unsolved problem of existence.

In the context of the bootstrap formfactor program for factorizing models one
observes an unexpected (and may be even undeserved) simplicity in the analytic
dependence of the formfactors on the coupling strength. There is always a
region around zero in which the coupling dependence is analytic. According to
general structural arguments there is however no reason that the yet unknown
correlation functions will inherit this property. This raises the general
question: do on-shell observables have better analytic properties in the
coupling than off-shell operators?

It is very interesting to compare the constructive control one has on the
basis of the Osterwalder-Schrader setting with that for models constructed in
modular bootstrap formfactor program. In the first case the restriction comes
from short distance properties; in the almost 40 years history it has not been
possible to go beyond superrenormalizable models (mainly $P\phi_{2}$). On the
other hand all the known factorizing models have strictly renormalizable
interactions (e.g. the sine-Gordon model interaction is nonpolynomial) and
there is no overlap. The weakness of one construction method is the strength
of the other. If one could break the limitation set by factorizability as
indicated above then the constructive approach would change in favor of the
modular wedge generator approach.

Factorizing models are very closely related to chiral conformal theories which
``live'' (in the sense of modular localization) on a (compactified) lightray.
On the one hand there is the general relation of a QFT to its scale invariant
short distance limit. Many different massive theories have the same critical
limit i.e. belong to one \textit{short distance universality class}. If one
only looks at factorizing models than Zamolodchikov has presented conditions
under which one can invert this relation in a formal setting of a perturbed
conformal theory \cite{pert}. On the other hand there is a conceptually quite
different relation between d=1+1 massive theories to their chiral
\textit{holographic projection \cite{2-dim}}. In that relation the algebraic
substrate and the Hilbert space in which it is represented remains unchanged
and only the spacetime indexing of the algebras is radically changed in a way
that cannot be encoded in a simple geometric relation between the chiral
fields on the lightray. But different from the AdS-CFT holography, the
\textit{lightfront holography is also a class property} i.e. without
enlargement of the Hilbert space there are many ambient theories which are
holographic inverses. Only if one had the luck to find generators of the
holographic projection which are covariant under the ambient Poincar\'{e}
group, as it is the case with the Z-F generators in factorizing models, the
holographic inverse is uniquely fixed. The rather complicated connection
between pointlike generators of the ambient algebra and those of its
holographic projection prevent an understanding of this relation by a
straightforward inspection.

\section{A modular analog of O-S setting and of the Nelson-Symanzik duality}

In those cases where the Schwinger functions associated with the O-S
Euclideanization admits a stochastic interpretation in the sense of Nelson and
Symanzik, one observes a very strong analogy to the modular localization as
will be explained in the following.

The issue of understanding Euclideanization in chiral theories became
particularly pressing after it was realized that Verlinde's
observation\footnote{Verlinde discovered a deep connection between fusion
rules and modular transformation properties of characters of rational
irreducible representations of chiral observable algebras.} is best understood
by making it part of a wider investigation involving angular parametrized
thermal n-point correlation functions of observable fields $\Phi_{i}$ in the
superselection sector $\rho_{\alpha}$
\begin{align}
\left\langle \Phi(\varphi_{1},..\varphi_{n})\right\rangle _{\rho_{\alpha}%
,2\pi\beta_{t}}  &  :=tr_{H_{\rho_{\alpha}}}e^{-2\pi\beta_{t}\left(
L_{0}^{\rho_{\alpha}}-\frac{c}{24}\right)  }\pi_{\rho_{\alpha}}(\Phi
(\varphi_{1},..\varphi_{n}))\,\\
\Phi(\varphi_{1},..\varphi_{n})  &  =\prod_{i=1}^{n}\Phi_{i}(\varphi
_{i})\nonumber\\
\left\langle \Phi(\varphi_{1},..\varphi_{n})\right\rangle _{\rho_{\alpha}%
,2\pi\beta_{t}}  &  =\left\langle \Phi(\varphi_{n}+2\pi i\beta_{t},\varphi
_{1},..\varphi_{n-1})\right\rangle _{\rho_{\alpha},2\pi\beta_{t}}\nonumber
\end{align}
where the first line defines the angular thermal correlations in terms of a
$L_{0}$- Gibbs trace at inverse temperature $\beta=2\pi\beta_{t}$ on
observable fields in the representation $\pi_{\rho_{\alpha}}.$ Gibbs states
are special (unnormalized) KMS states i.e.states whose correlations fulfill
the analytic property in the third line. Their zero point function which is
the Gibbs trace of the identity, defines the $L_{0}$- partition functions. In
contrast to the previously used ground states such thermal correlations are
\textit{independent on the particular localization of charges} $loc\rho
_{\alpha}.$ This is the result of the unitary invariance of the trace and
consequently they only depend on the equivalence class i.e. on the sector
$\left[  \rho_{\alpha}\right]  \equiv\alpha,$ which makes them valuable
objects to study the sector structure (classes of inequivalent representations
of the observable algebra). These correlation functions\footnote{The conformal
invariance actually allows a generalization to complex Gibbs parameters $\tau$
with $Im\tau=\beta$ which is however not neede in the context of the present
discussion.} fulfill the following amazing thermal duality relation
\begin{align}
\left\langle \Phi(\varphi_{1},..\varphi_{n})\right\rangle _{\alpha,2\pi
\beta_{t}}  &  =\left(  \frac{i}{\beta_{t}}\right)  ^{a}\sum_{\gamma}%
S_{\alpha\gamma}\left\langle \Phi(\frac{i}{\beta_{t}}\varphi_{1},..\frac
{i}{\beta_{t}}\varphi_{n})\right\rangle _{\gamma,\frac{2\pi}{\beta_{t}}%
}\label{duality}\\
a  &  =\sum_{i}dim\Phi_{i}\nonumber
\end{align}
where the right hand side formally is a sum over thermal expectation at the
inverse temperature $\frac{2\pi}{\beta_{t}}$ at the analytically continued
pure imaginary angles scaled with the factor $\frac{1}{\beta_{t}}.$ The
multiplicative scaling factor in front which depends on the scaling dimensions
of the fields $\Phi_{i}$ is just the one which one would write if the
transformation $\varphi\rightarrow\frac{i}{\beta_{t}}\varphi$ were an ordinary
conformal transformation law. Before presenting a structural derivation of
this relation the reader should notice the analogy with the thermal version of
the Nelson-Symanzik for massive two-dimensional theories (second section).
Since chiral theories are localized on the (compactified) lightray, the analog
of the Euclidean spacetime interchange consists of an interchange of the angle
with its imaginary version; the stretching factor $\frac{1}{\beta_{t}}$
together with the inverse temperature corresponds to the interchange of the
two periodicities in the N-S duality. The appearance of the linear combination
of all (finite in rational models) superselection sectors weighted with the
Verlinde matrix $S$ has no counterpart in the N-S setting. In simple models as
e.g. the multi-component abelian current model \cite{2-dim} the proof of the
temperature duality relation can be reduced to properties of the Dedekind
eta-function, the Jacobi $\Theta$-functions as well as the Poisson-resummation
property. The Kac-Peterson-like character relations%
\begin{equation}
\chi_{\alpha}(\tau)=\sum_{\beta}S_{\alpha\beta}\chi_{\beta}(-\frac{1}{\tau
}),\,\,\chi_{\alpha}(\tau)\equiv tr_{H_{\rho_{\alpha}}}e^{-2\pi\beta
_{t}\left(  L_{0}^{\rho_{\alpha}}-\frac{c}{24}\right)  }\mathbf{1}%
\end{equation}
is a special case of the (\ref{duality}) when one uses instead of fields the
identity. The relevant Verlinde matrix $S$ is the one which diagonalizes the
$Z_{N}$ lattice fusion rules and together with a certain diagonal matrix $T$
generates a unitary representation of the modular group $SL(2,\mathbb{Z})$
whose generators are T: $\tau\rightarrow\tau+1$ and S: $\tau\rightarrow
-\frac{1}{\tau}$ But a profound understanding of its content can only be
achieved by a general structural argument. Under certain technical assumption
within the setting of vertex operators\footnote{The Vertex framework is based
on pointlike covariant objects, but unlike Wightman's formulation it is not
operator-algebraic i.e. the star operation is not inexorably linked to the
topology of the algebra as in $C^{\ast}$algebras of quantum mechanical origin.
Although it permits a generalization beyond two dimensions \cite{Nic}, the
determination of classifications and representations of higher-dimensional
vertex-algebras remains an open problem. \ }, Huang recently presented a
structural (model-independent) proof \cite{Huang} of the character relation
with Verlinde's definition of $S$. Huang's proof does not really reveal the
deep local quantum physical principles which the analogy to the N-S duality suggests.

The fact that the character relation is a special case of a relation which
involves analytic continuation to imaginary rotational lightray coordinates
suggests that one should look for a formulation in which the rotational
Euclideanization has a well-defined operator-algebraic meaning. On the level
of operators a positive imaginary rotation is related to the Moebius
transformation $\tilde{\Delta}^{it}$ with the two fixed points $(-1,1)$ via
the formula%
\begin{equation}
e^{-2\pi\tau L_{0}}=\Delta^{\frac{1}{4}}\tilde{\Delta}^{i\tau}\Delta
^{-\frac{1}{4}}=\tilde{\Delta}_{c}^{i\tau} \label{cont}%
\end{equation}
where $\Delta^{it}$ and $\tilde{\Delta}^{it}$ represents the $SL(2,R)$ Moebius
subgroups with fixpoints $(0,\infty)$ resp. $(-1,1)$ and $\tilde{\Delta}%
_{c}^{i\tau}$ the $SU(1,1)$ subgroup with $z=(e^{-i\frac{\pi}{2}}%
,e^{i\frac{\pi}{2}})=(-i,i)$ being fixed (the subscript $c$ denotes the
compact picture description). Note that $Ad\Delta^{\frac{1}{4}}$ acts the same
way on $\tilde{\Delta}^{i\tau}$ as the Cayley transformation $AdT_{c},$ where
the $T_{c}$ is the matrix which represents the fractional acting Cayley
transformation
\begin{equation}
T_{c}=\frac{1}{\sqrt{2}}\left(
\begin{array}
[c]{cc}%
i & 1\\
-i & 1
\end{array}
\right)
\end{equation}
Ignoring for the moment domain problems for $\Delta^{\frac{1}{4}}$, the
relation (\ref{cont}) gives an operator representation for the analytically
continued rotation with positive imaginary part $(t>0)$ in terms of a Moebius
transformation with real rapidity parameter. If we were to use this relation
in the vacuum representation for products of pointlike covariant fields $\Phi$
where the spectrum of $L_{0}$ is nonnegative, we would with obtain with
$\Phi(t)=e^{2\pi itL_{0}}\Phi(0)e^{-2\pi itL_{0}}$%
\begin{align}
\left\langle \Omega\left|  \Phi_{1}(it_{1})..\Phi_{n}(it_{n})\right|
\Omega\right\rangle _{{}}^{ang}  &  =\left\langle \Omega\left|  \Phi_{1}%
(t_{1})_{c}..\Phi_{n}(t_{n})_{c}\right|  \Omega\right\rangle ^{rap}%
\label{vacuum}\\
&  =\omega_{2\pi}(\Phi_{1}(t_{1})_{c}..\Phi_{n}(t_{n})_{c})^{rap}\nonumber
\end{align}
The left hand side contains the analytically continued rotational Wightman
functions. As a result of positivity of $L_{0}$ in the vacuum representation
this continuation is possible as long as the imaginary parts remain ordered
i.e. $\infty>t_{1}>...>t_{n}>0.$ On the right hand side the fields are at
their physical boundary values parametrized with the rapidities of the compact
$\tilde{\Delta}_{c}^{it}$ Moebius subgroup of $SU(1,1)$. Note that this
rapidity interpretation implies a restriction since the rapidities associated
with $x=th\frac{t}{2}$ cover only the interval $(-1,1).$\ The notation in the
second line indicates that this is a KMS state at modular temperature
$\beta_{mod}=1$ ($\beta_{Hawking}=2\pi\beta_{mod}=2\pi$) in agreement with the
well-known fact that the restriction of the global vacuum state to the
interval (-1,1) becomes a state at fixed Hawking-Unruh temperature $2\pi.$
Note that only the physical right hand side is a Wightman \textit{distribution
in terms of a standard }$i\varepsilon$\textit{ boundary prescription}, whereas
the left hand side is an \textit{analytic function (i.e. without any boundary
prescription)}. This significant conceptual (but numerical harmless)
difference is responsible for the fact that in the process of angular
Euclideanization of chiral models\textbf{ }\textit{the KMS
condition\footnote{Contrary to popular believes KMS is not equivalent to
periodicity in time but it leads to such a situation if the the involved
operators commute inside the correlation function (e.g. spacelike separated
observables). } passes to a periodicity property and vice versa.}

The analogy with the generalized Nelson-Symanzik situation suggests to start
from a rotational thermal representation in the chiral setting. For simplicity
let us first assume that our chiral theory is a model which possess besides
the vacuum sector no other positive energy representations. Examples are
lattice extension of multicomponent Weyl algebras with selfdual lattices (e.g.
the moonshine lattice). In this case $S=1$ in the above matrix relation
(\ref{duality}). Assume for the moment that the Gibbs temperature is the same
as the period namely $\beta_{mod}=1.$ According what was said about the
interchange of KMS with periodicity in the process of angular Euclideanization
we expect the selfdual relation
\begin{align}
&  \left\langle \Omega_{1}|\Phi(it_{1})...\Phi(it_{n})|\Omega_{1}\right\rangle
^{rot}=\left(  i\right)  ^{ndim\Phi}\left\langle \Omega_{1}^{E}\right|
\Phi^{E}(t_{1})...\Phi^{E}(t_{n})\left|  \Omega_{1}^{E}\right\rangle
^{rot}\label{an}\\
&  \left\langle \Omega_{1}|\Phi(t_{1})...\Phi(t_{n})|\Omega_{1}\right\rangle
^{rot}\equiv tr(\Omega_{1},\Phi(t_{1})...\Phi(t_{n})\Omega_{1}),\,\Omega
_{1}\equiv e^{-\pi L_{0}}\nonumber\\
&  \Phi^{E}(t_{1})^{\dagger}\equiv\tilde{J}\Phi^{E}(t_{1})\tilde{J}=\Phi
^{E}(-t_{1})^{\ast},\,\left[  \tilde{J},L_{0}\right]  =0\nonumber
\end{align}
where the analyticity according to a general theorem about thermal states
\cite{Kl-La}\cite{F} limits the $t^{\prime}s$ to the unit interval and
requires the ordering $1>t_{1}>...>t_{n}>0.\,$\ Thermal Gibbs states are
conveniently written in the Hilbert space inner product notation with the help
of the Hilbert-Schmidt operators $\Omega_{1}\equiv e^{-\pi L_{0}},$ in which
case the modular conjugation is the action of the Hermitian adjoint operators
from the right on $\Omega_{1}$ \cite{Haag}. Since the KMS and the periodicity
match crosswise, the only property to be checked is the positivity of the
right hand side i.e. that the correlations on the imaginary axis are
distributions which fulfill the Wightman positivity. Here the label $E$ on
$\Phi(t_{1})$ denotes the Euclideanization. For this we need the star
conjugation associated with $\tilde{J}$ which interchanges the right with the
left halfcircle which because of $L_{0}=H+\tilde{J}H\tilde{J}$ commutes with
$L_{0}.$ In that case the modular group of $\Phi^{E}(t_{1})=\Phi(it)$ is
$e^{-2\pi tL_{0}}$ and the modular conjugation is the Ad action of $\tilde{J}$
which changes the sign of $t$ as in the third line (\ref{an}). Whereas the
modular conjugation in the original theory maps a vector $A\Omega_{1}$ into
$\Omega_{1}A^{\ast}$ with the star being the Hermitean conjugate, the
Euclidean modular conjugation is $A^{E}\Omega_{1}^{E}\rightarrow\Omega_{1}%
^{E}\tilde{J}A^{E}\tilde{J}\equiv\Omega_{1}^{E}\left(  A^{E}\right)
^{\dagger}.$ This property is at the root of the curious selfconjugacy
(\ref{an}). .

There are two changes to be taken into consideration if one passes to a more
general situation. The extension to the case where one starts with a $\beta$
Gibbs state which corresponds in the Hilbert-Schmidt setting to $\Omega
_{\beta}=e^{-\pi\beta L_{0}}$ needs a simple rescaling $t\rightarrow\frac
{1}{\beta}t$ on the Euclidean side in order to maintain the crosswise
correspondence between KMS and periodicity. Since the Euclidean KMS property
has to match the unit periodicity on the left hand side, the Euclidean
temperature must also be $\frac{1}{\beta}$ i.e. the more general temperature
duality reads%
\begin{equation}
\left\langle \Omega_{\beta}|\Phi(it_{1})...\Phi(it_{n})|\Omega_{\beta
}\right\rangle ^{rot}=\left(  \frac{i}{\beta}\right)  ^{ndim\Phi}\left\langle
\Omega_{\frac{1}{\beta}}^{E}\right|  \Phi^{E}(t_{1})...\Phi^{E}(t_{n})\left|
\Omega_{\frac{1}{\beta}}^{E}\right\rangle ^{rot}%
\end{equation}
The positivity argument through change of the star-operation remains
unaffected. This relation between expectation values of pointlike covariant
fields should not be interpreted as an identity between operator algebras. As
already hinted at the end of section 2 one only can expect a sharing of the
analytic core of two different algebras whose different star-operations lead
to different closure. In particular the above relation does not represent a
symmetry in the usual sense.

The second generalization consists in passing to generic chiral models with
more superselection sectors than just the vacuum sector. As usual the systems
of interests will be rational i.e. the number of sectors is assumed to be
finite. In that case the mere matching between KMS and periodicity does not
suffice because all sectors are periodic as well as KMS and one does not know
which sectors to match. A closer examination (at the operator level taking the
Connes cocycle properties versus charge transportation around the circle into
account) reveals that the statistics character matrix $S$ \cite{Char}\ enters
as in (\ref{duality}) as a consequence of the well-known connection between
the \textit{invariant} content (in agreement with the sector $\left[
\rho\right]  $ dependence of rotational Gibbs states) of the circular charge
transport and the statistics character matrix \cite{F-R-S2}\cite{Fr-Ga}. For
those known rational models for which Kac-Peterson characters have been
computed, this matrix $S$ turns out to be identical to the Verlinde matrix $S$
which diagonalizes the fusion rules \cite{Verl} and which together with a
diagonal phase matrix $T$ $\ $generates a unitary representation of the
modular group $SL(2,Z)\footnote{Whereas relativistic causality already leads
to an extension of the standard KMS $\beta$-strip analyticity domain to a
$\beta$-tube domain \cite{Br-Bu}, conformal invariance even permits a complex
extension of the temperature parameter to $\tau$ with $Im\tau>0.$ For this
reason the chiral theory in a thermal Gibbs state can be associated with a
torus in the sense of a Riemann surface, but note that in \textit{no physical
sense} of localization this theory lives \textit{on} a torus.}.$ Confronting
the previous zero temperature situation of angular Euclidean situation with
the asymptotic limit of the finite temperature identity, one obtains the
Kac-Wakimoto relations as an identity between the temperature zero limit and
the double limit of infinite temperature (the chaos state) and short distances
on the Euclidean side.

This superselection aspect of angular Euclideanization together with the
problem in what sense this modular group $SL(2,Z)$ can be called a new
symmetry is closely related to a more profound algebraic understanding of the
relation between the analytic cores of the two algebras and requires a more
thorough treatment which we hope to return to in a separate publication.

Modular operator theory is also expected to play an important role in bridging
the still existing gap between the Cardy \cite{Cardy} Euclidean boundary
setting and those in the recent real time operator algebra formulation by
Longo and Rehren \cite{Lo-Re}

\section{Open problems, concluding remarks}

The comparision of the constructive results obtained in the O-S setting and in
the old bootstrap-formfactor approach built on the Smirnov construction
recipes with the more recent constructions based on modular localization
theory gives rise to a wealth of unsolved basic problems of QFT. Here are some
of them.

\begin{itemize}
\item  The O-S formulation and the modular setting are related in a deep and
yet mostly unknown way. In order to learn something about this connection one
may start with the Wigner representation setting. Recently Guerra has spelled
out what the O-S Euclideanization means in the simplest context of the
spinless one-particle Wigner space \cite{Gu2}. On the other hand all problems
concerning the modular localization setting have been explicitly answered for
all positive energy representations \cite{M-S-Y}. It would be very interesting
to translate these results into the O-S setting.

\item  The old bootstrap dream remained unfulfilled beyond factorizing, and
the new modular setting not only explains why a general pure S-matrix approach
is not feasible but also indicates that if one views the S-matrix construction
as part of a wider framework which aims at generators of wedge localized
algebras, this dream still may find its realization in a new construction of
QFT which bypasses correlation functions of (necessarily singular) correlation
functions of pointlike generators. With new hindsight and a new conceptual
setting one should revisit the properly re-formulated old problems.

\item  The d=1+2 massive Wigner representation can have anomalous (not
semiinteger) spin which leads to plektonic (braid group) statistics. The
simplest abelian family of representations is that of $Z_{N}$-anyons. Such
representations activate representations of the Poincar\'{e} group in which
the Lorentz part is represented through the Bargmann covering $\widetilde
{SO(2,1)}.$ The modular theory of these string-like representations has been
worked out in \cite{Mu1} and it is known that d=1+2 anomalous spin
representations are the only Wigner representations whose associated QFT has
vacuum-polarization which prevent the standard on-shell free field realization
\cite{Mu2}. It would be very interesting to understand how an O-S like
Euclidean formulation would look like.

\item  Up to now models of QFT have been ``baptized'' and studied in the
setting of Lagrangian quantization (either canonical of functional integral).
More recently the bootstrap-formfactor setting led to models which do not
possess a Lagrangian description (e.g. the $Z_{N}$ model in \cite{B-F-K} whose
natural description is in terms of $Z_{N}$ braid group statistics). There are
indications that interactions in terms of string-localized fields (which
apparently do not permit a Euler-Lagrange characterization) extends the
possibilities for formulating interactions. The ghostfree potential for the
physical fields of zero mass finite helicity representations (e.g. the
vectorpotential associated with the electromagnetic field) are necessarily
string-localized \cite{M-S-Y}. Also in this case it should be possible to use
these objects outside the Lagrangian framework directly in the implementation
of interactions. Such a description would be particularily interesting for
higher helicities as in the case of the graviton. An intrinsic description of
QFT ``without the classical crutches'' of Lagrangian quantization is an old
dream of Pascual Jordan, the protagonist of ``Quantelung der Wellenfelder''.
The continuation of the ongoing attempts may still lead to a fulfillment of
this dream.

\item  The relation between heat bath thermal behavior and the purely quantum
thermal manifestations of vacuum polarization (Hawking, Unruh, Bekenstein)
have received a lot of recent attention, but they still have not been
adequately understood. Modular properties (especially the split property),
analytic continuation and Euclideanization are expected to play an important role.
\end{itemize}

Quantum field theory, which in the frame of mind of some string theorist has
become a historical footnote of their theory, had already been declared dead
on two previous occasions; first in the pre-renormalization ultraviolet crisis
of the 30s and then again by the protagonists of the S-matrix bootstrap in the
60s. But each time a strengthened rejuvenated QFT re-appeared. It is clear
that an area which still produces many fundamental new questions is very far
from its closure.

There is however a new sociological problem which poses a serious impediment
to the kind of physics as it developed over several centuries through rational
discourse and which does not necessarily aim at a ``theory of everything''.
Since the times of Galilei and Newton its aim was the de-mystification of
nature and in this role it had an enormous impact on the European
enlightenment and more general on western civilization. This is presently
threatened by a new trend of re-mystification. Whereas this is most visible in
the Kulturkampf which the movement of Intelligent Design in the US unleashed
against Darwin's formation of species, the trend of re-mystification has
already entered particle physics. When one of the most influential particle
physicists invites (without any irony) the physics community to interpret the
meaning of the big letter ``M'' in M-theory as ``Mystery'' the aims have
already been redefined and the new direction certainly is not of a critical
discourse. It is also plainly visible in recent outings by well-known string
theorists \cite{Suss} and the question of whether anthropic arguments are
camouflaged Intelligent Design arguments is somewhat academic as far as the
future of particle physics is concerned since hegemonic aspects of string
theory (``There is no other Game in Town'' \cite{crisis}) are even more
detrimental to the fundamental research in particle physics than the
imposition of religious beliefs which does not have a direct impact on the
substance of research.  

It is naive to believe that in times of globalization there can be any area of
human activities which can be kept protected from the intellectual and
material arrogance of the new Zeitgeist of the post cold war era. A
\textit{Hegemon}\ will not change the established terminology but he can and
does re-define its meaning. The concepts of human rights are not abandoned,
they are just redefined to suit the Hegemon.

The present crisis in particle physics is not an isolated passing event, it
has a solid material basis. The hegemonial tendency in physics does not favor
conceptual progress through the dialectic sharpening of contradictions and
antinomies with the existing principles with the aim to reach a breaking point
from where a new principle could take over. The market forces rather favor the
much faster path to personal fame by contributions which basically consists in
the acclamation of prevailing fashions. Through the hegemonial rein of the
market the possibility that somebody in an old-fashioned patent office will
have the productive leisure to follow his own innovative ideas is practically
ruled out; this only remains as a nostalgic picture. The modern role model of
a particle physics theoretician is rather that of a citation-supported young
star who tries to stay on top of fashions by following all updates of big
Latin Letters. He is always prepared to support any change by cranking out new
computations and leaving aside any conceptual confrontation with traditional
principles for which his training in any case would be insufficient. A novice
in QFT would endanger his career\footnote{A warning example for young
physicists that straying away too much from dominating fashions may wreck an
academic career is the fate of H.-W. Wiesbrock, who a short time after his
innovative work on the application of modular operator theory in particle
physics at the FU did not find any place to continue his career.} if instead
of the fast calculational entrance into one of the ongoing globalized fashions
he would choose the more arduous path of getting to the conceptual and
mathematical roots of a problem, including some at least rudimentary knowledge
about its history. This traditional path was still possible for the generation
which includes Karowski and Schrader (and myself); in those days there were
fashions, but they they did not yet grow into hegemonially managed
monocultures. Nowadays it is very easy to compile a list of ``who is who'' in
the administration of the particle physics crisis. One only has to look at the
list of the editorial board of recently founded journals, a particular valid
illustration is JHEP. With such powerful control over the globalized impact of
papers and the academic market it is clear that this is not going to be a
transitory phenomenon of short duration.

Sociologists and historians of physics have tried to analyze the amazing
progress which took place in war-torn Europe (in particular Germany) in the
aftermath of world war 1. Some have attributed the loss of certainty in favor
of probabilistic concepts in the emerging quantum theory to the gloom and doom
Zeitgeist \cite{For} (which found its expression in the widely red historical
treatise ``The Decline of the West'' by Oswald Spengler) from where, so they
argue, one could become part of the avant-garde only by a very revolutionary
tabu-breaking conceptual step. Such explanations do not appear very plausible.
It should be considerably more natural to explain the present crisis in terms
of all-pervading rule of the globalized market in which impact parameter and
personal fame (and not the gain of genuine knowledge) are the propelling forces.

\textit{Acknowledgements}: I am indebted to Walter Wreszinski for several
clarifications and suggestions. I also thank the members of the Fisica
Matematica department at the USP (where most of this work was carried out) for hospitality.

\end{document}